\documentclass{emulateapj}
\usepackage{amsmath,amsbsy}

\begin{document}

\title{Microwave-Sky Simulations and Projections for Galaxy Cluster Detection with the Atacama Cosmology Telescope}

\author{Neelima Sehgal}
\affil{Department of Physics and Astronomy, Rutgers University, 136
Frelinghuysen Road, Piscataway, NJ 08854}
\author{Hy Trac}
\affil{Department of Astrophysical Sciences, Peyton Hall, Princeton
University, Princeton, NJ 08544}
\author{Kevin Huffenberger}
\affil{Jet Propulsion Laboratory, 4800 Oak Grove Drive, Pasadena, CA
91109\\ California Institute of Technology, Pasadena, CA 91125}
\author{Paul Bode}
\affil{Department of Astrophysical Sciences, Peyton Hall, Princeton
University, Princeton, NJ 08544}

\begin{abstract}
We study the ability of three-frequency, arcminute-resolution
microwave measurements to detect galaxy clusters via their
Sunyaev-Zel'dovich (SZ) distortion of the microwave background.  For
this purpose, we have constructed large-area simulations of the
microwave sky, and we have made them publicly available to further
investigations into optimal data reduction techniques for upcoming
SZ cluster surveys. In these sky simulations, galaxy clusters are
modeled using N-body simulated dark matter halos plus a gas
prescription for the intracluster medium that allows the small scale
cluster physics such as star formation and feedback to be
realistically incorporated. We also model the primary microwave
background, radio and infrared point sources, galactic dust
emission, and the SZ flux including kinetic and relativistic
contributions. We make use of these simulations to study the
scaling relation between integrated SZ flux and cluster mass, and
find our clusters follow a power-law with an index that
is steeper than that for self-similar cluster models. Some evolution
of the power-law index and normalization with redshift is also
observed. These simulations are also used to study cluster detection
for the Atacama Cosmology Telescope (ACT). Using a multi-frequency
Wiener filter to separate clusters from other microwave components,
we find that ACT alone can recover a cluster sample that is
$\approx$ 90\% complete and $\approx$ 85\% pure
above $3 \times 10^{14} M_{\odot}$.
\end{abstract}


\section{INTRODUCTION}

Cluster catalogs provide valuable information on the evolution and
distribution of matter over cosmic time.  It has long been realized
that the abundance of galaxy clusters as a function of redshift is a
sensitive probe of the underlying cosmology
\citep[e.g.][]{Oukbir1992, Eke1996, Viana1996, Barbosa1996,
Bahcall1998, Rosati2002}. It has also been appreciated that a large
sample of galaxy clusters can provide important constraints on the
dark energy density and equation of state which are complementary to
those obtained from microwave background and type Ia supernovae
measurements \citep[e.g.][]{Wang1998, Haiman2001, Weller2002,
Weller2003, Majumdar2003, Wang2004, Lima2005}.  This has prompted
efforts to obtain cluster catalogs from wide, deep surveys.

Searches for clusters based on their Sunyaev-Zel'dovich (SZ) effect
\citep{Sunyaev1970,Sunyaev1972} have a particular advantage over
X-ray and optical searches since the SZ signal does not dim at high
redshifts (z$>$1), where the cluster abundance is strongly dependent
on cosmology.  (For reviews on the SZ effect see
\citet{Birkinshaw1999} and \citet{Carlstrom2002}).  In light of
this, there are a host of ground-based bolometer array instruments
(ACBAR, ACT, APEX, SPT) and interferometers (AMI, AMiBA, SZA) either
online or scheduled to come online in the next few years that
potentially will detect hundreds to thousands of clusters at
microwave frequencies via their SZ effect
\citep{Runyan2003,Kosowsky2003,Gusten2006,Ruhl2004,Kneissl2001,Li2006}.
The \emph{Planck} satellite, to be launched in 2008, should also
provide an all-sky map of massive clusters in the microwave band
\citep{Tauber2004}.

Using cluster catalogs to constrain cosmology requires a solid
understanding of the cluster selection criteria.  A given survey's
cluster selection function is intimately tied to characteristics of
the instrument and data reduction techniques.  Considerable
attention has been devoted to exploring various data reduction
methods for different SZ surveys.  This has been pursued in an
effort to study both the optimal detection of clusters and the
optimal recovery of the cluster SZ flux.  The former consists of
maximizing the completeness and minimizing the contamination of a
given cluster sample.  The latter is necessary to properly identify
the observable (SZ flux) with mass, or to employ self-calibration
schemes to fit for the SZ flux - mass relation and cosmology
simultaneously \citep{Majumdar2004,Younger2006}.

Various filtering approaches have been utilized to isolate the SZ
signal and extract clusters in simulations.  Such techniques have
employed matched filters \citep{Herranz2002a,Herranz2002}, Wiener
filters \citep{Aghanim1997}, wavelet filters \citep{Pierpaoli2005},
and maximum entropy methods \citep{Stolyarov2002}.  Several papers
have implemented versions of these filters and begun careful study
of selection functions for upcoming SZ surveys
\citep{Schulz2003,White2003,Melin2005,Vale2006,Melin2006,Schaefer2006,Juin2007}.
With this in mind, we have created a large-scale simulation of the
microwave sky on which these data reduction techniques can be
refined.

To model the SZ flux accurately generally requires expensive,
high-resolution hydrodynamic simulations to realistically model the
small scale cluster physics.  However, it is a challenge to create
such simulations in a large enough volume so that one is not limited
by cosmic variance when performing statistical studies. To overcome
this challenge, we create a cluster catalog using an N-body
simulation combined with a gas prescription given by
\citet{Ostriker2005}. The N-body code only needs to be run once for
a given cosmology, and the small scale cluster physics (including
non-spherically symmetric gas distributions, star formation, and
feedback) are added afterward via the gas prescription.  As a
result, the cluster physics can be varied easily without having to
redo expensive runs.

These cluster simulations were incorporated within microwave-sky
simulations consisting of two strips of the sky, each 4 degrees wide
in declination and 360 degrees around in right ascension.  One strip
is centered at a declination of -5 degrees, and the other is
centered at a declination of -55 degrees, to match the portion of
sky that the Atacama Cosmology Telescope (ACT) will observe.  The
galaxy clusters within this simulation have halo masses down to $5
\times 10^{13}$ M$_{\odot}$, and the N-body simulation employs the
latest cosmological parameters derived from a combination of WMAP3,
SDSS, HST, and SN Astier observations ($\Omega_{m}$=0.26,
$\Omega_{\Lambda}$=0.74, $\Omega_{b}$=0.044, n=0.95,
$\sigma_{8}$=0.77, h=0.72) \citep[see][and references
therein]{Spergel2006}.  We include the full SZ effect, with
relativistic corrections, as well as infrared and radio point
sources (uncorrelated with clusters), using number counts given by
\citet{Borys2003} (infrared) and \citet{Knox2004} (radio). Galactic
dust and primary microwave background fluctuations are also modeled,
the later generated using the WMAP ILC map \citep{Bennett2003b,
Hinshaw2006}. These sky simulations were made at the observing
frequencies of 145, 215, and 280 GHz and are in a cylindrical equal
area sky projection \citep{Calabretta2002}.  The final sky maps are
in a standard FITS format and have a pixel size of $\approx
0.2'\times 0.2'$ and units of Jy/ster. A catalog of all the cluster
halos in the simulation is also provided, specifying each halo's
basic properties.  These microwave-sky simulations are available at
http://www.astro.princeton.edu/$\sim$act/.

We use these simulations to study the scaling relation between SZ
flux and cluster mass, as well as prospects for cluster detection
with ACT. Roughly $10^5$ clusters  are used to determine the
simulated SZ flux - mass scaling relation. We compare cluster
M$_{200}$ to Y$_{200}$, which we take to be the SZ Compton-y
parameter integrated over a disk of radius R$_{200}$. These clusters
are fit to a power-law relation between Y$_{200}\times E(z)^{-2/3}$
and M$_{200}$, and estimates of the power-law index and
normalization are given.

We also employ a multi-frequency Wiener filter and simple
peak-finding algorithm to forecast cluster detection given ACT
instrument specifications.  Since there has been considerable
interest of late in how point source contamination affects cluster
detection (e.g. \citet{Melin2006}), this is investigated under three
different contamination assumptions: no point sources, only infrared
point sources, and both infrared and radio point sources.
Completeness of our recovered cluster sample is given as both a
function of M$_{200}$ and Y$_{200}$ (in units of arcmin$^{2} $), the
latter being directly obtainable from microwave observations.  We
also give the purity of our projected cluster sample, where purity
is one minus the percentage of false-positive detections.  The issue
of optimal SZ flux recovery for individual clusters is left to later
work.

The outline of this paper is as follows.  In \S 2, we give the
simulation details. \S 3 consists of an investigation of the Y-M
relation as suggested by our simulations, and \S 4 gives cluster
detection projections for ACT. In \S 5, we discuss directions for
future work, and, in \S 6, we summarize and conclude.

\section{SIMULATIONS}

\subsection{Simulated Clusters}   \label{sec:simclust}

\subsubsection{The Dark Matter Run}   \label{sec:dmrun}

Initial conditions for the $N$-body run were generated with the
GRAFIC2 code \citep{2001ApJS..137....1B}, available at
\url{http://web.mit.edu/edbert/}. As written, this code uses a
spherical Hanning filter on small scales.  However, we have found
that this filter significantly suppresses power on these scales, so
we removed it from use. GRAFIC2 perturbs particles from a regular
grid using the Zel'dovich approximation.  The simulation started at
redshift z=35.3, when the initial density fluctuation amplitude on
the scale of this grid was 10\%.   $N=1024^3\approx 10^9$ particles
were contained in a periodic box of size $L=1000h^{-1}$Mpc, so the
particle mass is thus $6.72\times 10^{10}h^{-1}M_\odot$. The cubic
spline softening length was set to $\epsilon=16.276h^{-1}$kpc. The
simulation was carried out with the TPM code
\citep{2000ApJS..128..561B, 2003ApJS..145....1B}, modified slightly
from the publicly available version (at
\url{http://www.astro.princeton.edu/$\sim$bode/TPM/}). Particle
positions and velocities were followed at double precision, though
acceleration was kept at single precision.  Also, no lower limit was
set to the parameter $B$ used in domain decomposition \citep[see
Eqn. 5 of][]{2003ApJS..145....1B}, which at late times results in
more particles being followed at full force resolution.   The
initial domain decomposition parameters used are $A=1.9$ and
$B=9.2$. The PM mesh contained $2048^3$ cells, and the maximum
sub-box was 256 cells. By the end of the run, $5\times 10^6$ trees
containing 54\% of the particles were followed at full resolution.
More details of the simulation are in \citet{Bode2006}.

At each PM time step, the matter distribution in a thin shell is
saved. The radius of the shell corresponds to the light travel time
from a $z=0$ observer sitting at the origin of the box, and its
width corresponds to the time step interval. The portion of a
spherical shell covering one octant of the sky was saved, so for
comoving distances larger than $1000h^{-1}$Mpc there can be some
duplication in structures as the periodic box is repeated. However,
while there are repeated dark matter halos in the simulations, the
repetition is usually at a different redshift, so the dynamical
state is different.  In cases where a cluster appears twice at the
same redshift, it is viewed at two different angles, making each
projected image unique.  At the end of the simulation, the full
matter distribution in a light cone extending to $z=3$ was saved in
417 time slices. Dark matter halos in this light cone were
identified with the Friends-of-Friends (FoF) algorithm using a
linking parameter $b=0.2$ (i.e. the linking length is one fifth of
the mean interparticle separation of the simulation, or
195$h^{-1}$kpc comoving). A $5\times 10^{13}h^{-1}M_\odot$ cluster
contains 744 particles.

\begin{figure}
\begin{center}
\epsscale{1.17}\plotone{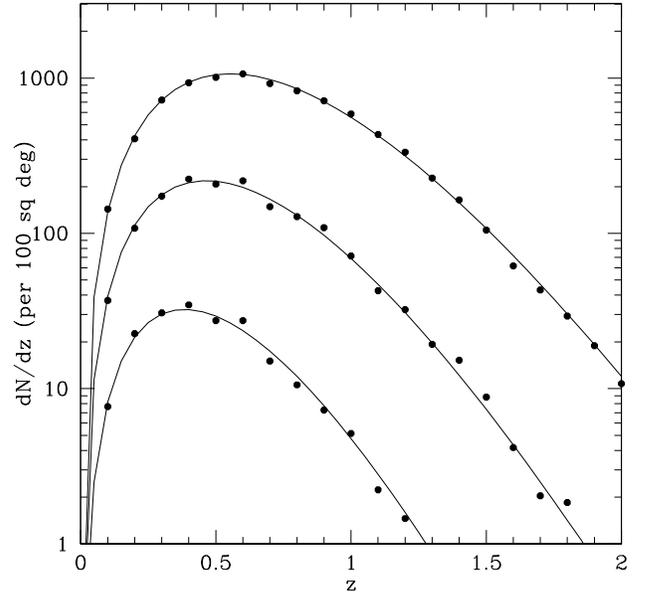} \caption{The halo abundance above a
minimum mass per unit redshift per 100 square degrees measured from
the simulation is compared with the semi-analytic fitting formula
from \citet{Jenkins2001}.  From top to bottom, the three sets of
points are for the minimum masses $M_{\rm min}=(1,2,4)\times
10^{14}M_\odot/h$.} \label{fig:dNdz}
\end{center}
\end{figure}

As shown in Figure \ref{fig:dNdz}, the halo mass functions measured
from the simulated light cone agree well with the semi-analytic
fitting formula from \citet{Jenkins2001} for a FoF linking length
$b=0.2$.  The number of halos above a minimum mass per unit redshift
per 100 square degrees is plotted for three minimum masses $M_{\rm
min}=(1,2,4)\times 10^{14}M_\odot/h$.  The data points are measured
using all halos in the octant to minimize sample variance, but then
are normalized to 100 square degrees.  For any cluster survey, the
minimum detectable mass is likely to be a function of redshift, and
the three values chosen reflect the range expected for the upcoming
SZ surveys.

\subsubsection{Adding Gas to Dark Matter Halos}   \label{sec:addgas}

Gas is added to each cluster following the prescription of
\citet{Ostriker2005} and \citet{Bode2006}. For each dark matter
halo, the gravitational potential is found on a mesh with cell size
$2\epsilon=32.552h^{-1}$kpc. The gas density and pressure (or
temperature) are found in each cell by assuming it is in hydrostatic
equilibrium and has a polytropic equation of state with adiabatic
index $\Gamma=1.2$. It is also assumed that initially the baryon
fraction inside the virial radius is $\Omega_b/\Omega_m$, and that
the energy per unit mass of the baryons equals that of the dark
matter. To specify the central density and pressure with these
assumptions requires two constraints.

The first constraint is the pressure at the virial radius, which is
the radius enclosing the virial overdensity calculated from
spherical tophat collapse \citep[this is a slight change from][which
used an overdensity of 200 times critical]{Ostriker2005}. The dark
matter kinetic energy is found in a buffer region extending for nine
cells outside of the virial radius. This is then translated into a
surface pressure, $P_s$,  by assuming the velocity dispersion is
isotropic and the baryon fraction in this region equals the cosmic
mean. The second constraint is conservation of energy.  The final
energy of the gas must equal this initial energy, after adjusting
for any of the effects that can cause changes in energy.

One such effect is that the gas distribution inside the cluster can
expand or contract.  This can, for example, reduce the baryon
fraction inside the virial radius if the gas distribution expands
further out. This would also reduce the energy of the gas as it does
mechanical work pushing against the surface pressure. The surface
pressure is assumed to remain at $P_s$ wherever the final gas radius
ends up.

Star formation will also affect gas energy.  At $z=0$, it is assumed
that the stellar to gas mass ratio is 0.10. The star formation rate
is assumed to follow a delayed exponential model \citep[Eqn. 1
of][]{Nagamine2006} with decay time $\tau=1$ Gyr, so at higher
redshift the star/gas mass ratio can be calculated. To make stars,
the gas with the largest initial binding energy is removed, thus
changing the total energy budget.  Furthermore, some fraction of the
rest mass turned into stars is taken to be converted into thermal
energy in the gas, via supernovae and AGN. The energy added to the
intracluster medium by these processes can be written as
$\epsilon_{f}M_*c^2$, where $M_*$ is the stellar mass and the
feedback efficiency is estimated to be $\epsilon_{f}=5\times
10^{-5}$. This value is determined by fitting to X-ray observations
of nearby clusters \citep{Bode2006}; at z=0, this added energy
amounts to roughly 3 keV per particle.

\begin{figure}
\begin{center}
\epsscale{1.17}\plotone{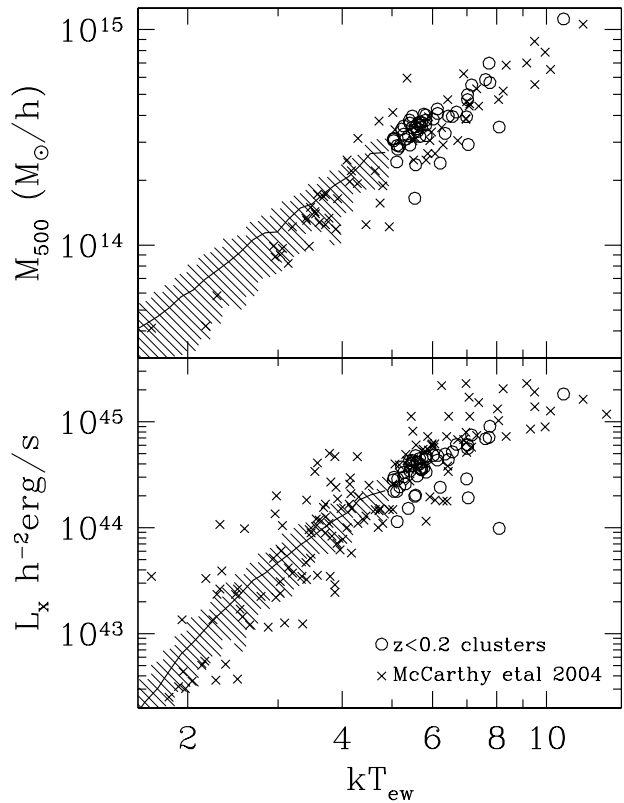} \caption{The panels show the
$M_{500}-T$ and $L_X-T$ relations for the simulated clusters below
$z=0.2$ compared to those from observed X-ray clusters.  The open
circles are the simulated clusters above $kT_{\mathrm{ew}}=5$keV.
The median value of the simulated clusters below 5 keV is
represented by a solid line, with the shaded region enclosing 90\%
of the clusters. $T_{\mathrm{ew}}$ is the emission-weighted X-ray
temperature.}\label{fig:XLandMvT}
\end{center}
\end{figure}

The upper panel of Figure \ref{fig:XLandMvT} shows the resulting
$M_{500}-T$ relation for all the clusters in the light cone at low
redshift, $z<0.2$. The temperature used is the X-ray
emission-weighted temperature inside $R_{500}$. The $kT>5$keV
clusters are shown as open circles, and, for clarity, the median
value of the clusters below 5 keV is shown as a line, with the
shaded region enclosing 90\% of the clusters. Shown for comparison
is data derived from X-ray clusters over the same redshift range,
taken from \citet{ReipBohr02} as adjusted by \citet{MBBPH04}.  The
simulated sample reproduces the observed $M_{500}-T$ relation quite
well. The gas temperature (or pressure) is fairly insensitive to the
exact choice of $\epsilon_{f}$.  Feedback has much more of an effect
on gas density, making the $L_X-T$ relation a more exacting
comparison \citep{Bode2006}. This relation is shown in the bottom
panel of Figure \ref{fig:XLandMvT}, along with data points taken
from the ASCA cluster catalog \citep{Horner01} as described in
\citet{MBBPH04}. Again, the simulated clusters agree with the
observed relation.

\subsection{Simulated SZ Signal}

For each of the simulated clusters, the gas prescription gives us
the electron number density $n_e$, temperature $T_e$, and velocity
$v$ fields with which to model the SZ effect, including relativistic
corrections.  We assume that the gas is fully ionized with a helium
mass fraction equal to 0.24.  The change in the microwave background
intensity after passing through a path length d$l$ in the direction
$\hat{n}$ is given by
\begin{align}
\frac{\Delta I_{\nu}}{I_{0}} & =
\frac{X^{4}e^{X}}{(e^{X}-1)^{2}}{\rm d}\tau \theta_{e} \bigl[Y_{0} +
\theta_{e} Y_{1} + \theta_{e}^{2} Y_{2} + \theta_{e}^{3} Y_{3} +
\theta_{e}^{4} Y_{4} \bigr]
                                         \nonumber\\
& {}+\frac{X^{4}e^{X}}{(e^{X}-1)^{2}}{\rm d}\tau (\frac{v}{c})^{2}
\bigl[\frac{1}{3} Y_{0} + \theta_{e} (\frac{5}{6} Y_{0} +
\frac{2}{3} Y_{1}) \bigr]
   \nonumber\\
& {}+\frac{X^{4}e^{X}}{(e^{X}-1)^{2}}{\rm d}\tau \frac{v_{\rm
los}}{c} \bigl[1 + \theta_{e} C_{1} + \theta_{e}^{2} C_{2} \bigr] ,
\label{eqn:SZ}
\end{align}
where
\begin{align}
I_{0}   &\equiv 2(k_{B} T_{CMB})^{3}/(hc)^{2} ,\\
X & \equiv h\nu/k_{B} T_{CMB} ,\\
\theta_{e} & \equiv k_{B}T_{e}/m_{e}c^{2} ,
\end{align}
and the $Y$'s and $C$'s are known frequency-dependent coefficients
\citep{Nozawa1998}.  The usual Compton y-parameter is given by
\begin{equation}
y \equiv \int\theta_{e}{\rm d}\tau = \frac{k_B \sigma_T}{m_e
c^2}\int n_e T_e {\rm d}l , \label{eqn:Compton-y}
\end{equation}
where ${\rm d}\tau = \sigma_{T} n_{e}{\rm d}l$ is the optical depth
through a path length ${\rm d}l$.  The first-order
(non-relativistic) thermal and kinetic SZ signals are given by
\begin{equation}
\left(\frac{\Delta T}{T_{\mathrm{CMB}}}\right)_{\mathrm{tsz}} \equiv
y Y_{0} = y \left(X{\rm coth}(X/2)-4\right) \label{eqn:thsz}
\end{equation}
and
\begin{equation}
\left(\frac{\Delta T}{T_{CMB}}\right)_{\mathrm{ksz}} \equiv
\int\left(\frac{v_{\rm los}}{c}\right){\rm d}\tau = \sigma_T\int n_e
\left(\frac{v_{\rm los}}{c}\right){\rm d}l ,
\end{equation}
respectively, where $v_{\rm los}$ is the line-of-sight component of
the peculiar velocity field.  For a 10 keV cluster with $v_{\rm
los}=1000$ km/sec, the correction to the first-order thermal SZ from
the $O$(${\rm d}\tau \theta_{e}^2$) term at 145 GHz is about 7.5\%,
but it becomes considerably more substantial near the null of the
thermal SZ.  The terms $O$(${\rm d}\tau v_{\rm los} \theta_{e})$ and
$O$(${\rm d}\tau v_{\rm los} \theta_{e}^2)$ give about 8\% and 1\%
corrections to the first-order kinetic SZ \citep{Nozawa1998}. Note
that the factor $X^4e^X/(e^X-1)^2$ in equation \ref{eqn:SZ} converts
between $\Delta T/T_{CMB}$ and $\Delta I_\nu/I_0$.

\begin{figure*}
\begin{center}
\epsscale{1.14}\plotone{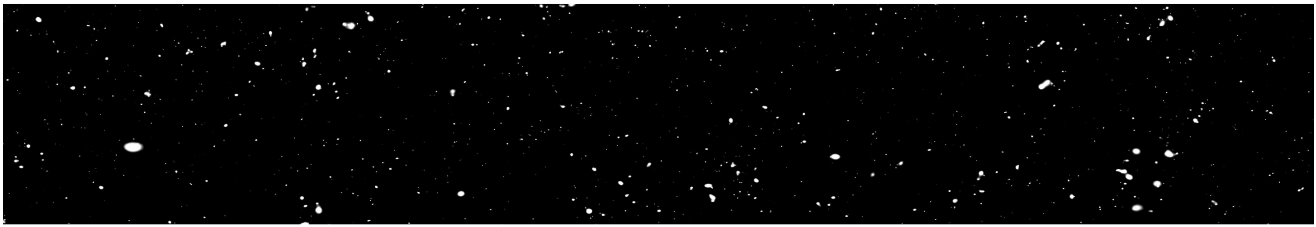} \caption{A sample image showing the
fully relativistic SZ signal at 145 GHz. The image dimensions are 33
degrees across in right ascension by 4 degrees wide in declination
centered at $\delta=-55$ degrees.  Black represents $\Delta
T/T_{CMB}$ greater than $-6 \times 10^{-6}$, and white represents
$\Delta T/T_{CMB}$ less than $-10 \times 10^{-6}$. The former value
is chosen to be representative of ACT instrument noise in 0.2'
pixels.} \label{fig:thsz_image}
\end{center}
\end{figure*}

Sky maps at the ACT observing frequencies of 145, 215, and 280 GHz
are made by tracing through the clusters and projecting them onto a
cylindrical equal-area grid.  Two strips of the sky, each 4 degrees
wide in declination and 360 degrees around in right ascension, are
constructed and centered at $\delta=-5$ and $\delta=-55$ degrees,
respectively.  Since the simulation light cone covers an octant,
only a quarter of the strip is unique and the other three-quarters
are mirrored.  The accompanying cluster catalog contains the
following information:  redshift, right ascension, declination,
$M_{\mathrm{FoF}}$, $M_{200}$, $M_{500}$, $R_{200}$, $R_{500}$, and
integrated SZ and X-ray properties. ($M_{\mathrm{FoF}}$ is the FoF
derived mass and $M_{200}$ and $M_{500}$ are the cluster masses
within $R_{200}$ and $R_{500}$ respectively, where the latter are
the radii at which the cluster mean density is 200 and 500 times the
critical density at the cluster redshift.)  Figure
\ref{fig:thsz_image} is a sample image showing the fully
relativistic SZ signal at 145 GHz. This image is 33 degrees across
in right ascension and 4 degrees wide in declination, centered at
$\delta=-55$ degrees.  Black represents $\Delta T/T_{CMB}$ greater
than $-6 \times 10^{-6}$, and white represents $\Delta T/T_{CMB}$
less than $-10 \times 10^{-6}$. The former value is chosen to be
representative of ACT instrument noise in 0.2' pixels.

\begin{figure}
\begin{center}
\epsscale{1.1}\plotone{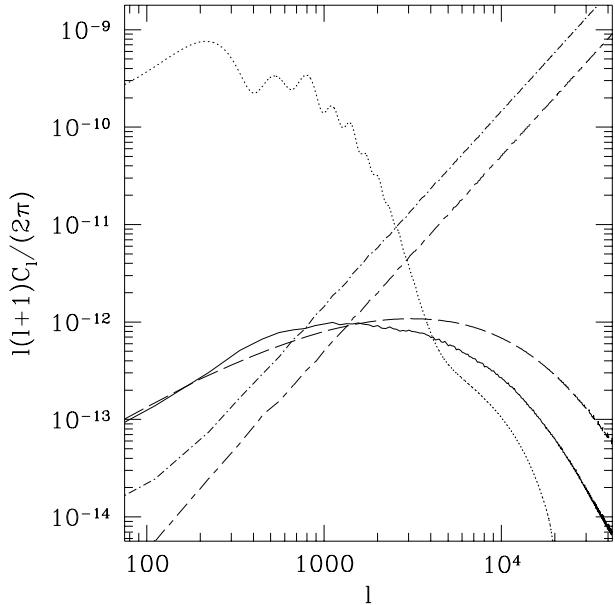} \caption{Solid curve is the
first-order thermal SZ power spectrum at 145 GHz from the
simulation.  Dashed curve is the first-order thermal SZ power
spectrum at 145 GHz derived analytically from \citet{Komatsu2002}.
For comparison, we include the lensed primary microwave background
power spectrum derived from CAMB \citep{Challinor2005} for our input
cosmology (dotted) and the power spectrum of infrared (dot-dashed)
and radio (dash-dashed) point sources from the simulation, also at
145 GHz. The radio source power spectrum includes only sources
fainter than 35 mJy. Note that $C_l$ is in dimensionless units of
$(\Delta T/T)^2$.} \label{fig:thsz}
\end{center}
\end{figure}

In Figure \ref{fig:thsz}, the first-order thermal SZ power spectrum
from the simulated map at 145 GHz is compared with a semi-analytic
prediction derived following the prescription of
\citet{Komatsu2002}.  On large angular scales, $l\sim100$, the two
spectra are in agreement, but deviations on smaller scales are found.
This is to be expected,  since
our gas prescription accounts for star formation and
feedback which reduce the gas density within clusters
\citep{Ostriker2005, Bode2006}, resulting in a suppression of power
on small ($l\gtrsim 1000$) scales.  In addition, the feedback pushes
the gas out to larger radii and increases the overall temperature of
the cluster, leading to an enhancement of power on scales $l\lesssim
1000$.  This demonstrates that
one must have a good understanding of baryonic
physics within clusters in order to extract cosmology from the SZ
power spectrum. Conversely, if the cosmology is well determined,
then one can learn about cluster physics from the detected SZ
signal.

\subsection{Simulated Primary CMB, Point Sources, and Galactic Dust}

A primary microwave background map is constructed to have the same
large scale structure as we observe on the sky and small scale
structure consistent with theoretical expectations.  For our large
scale map, we use the WMAP ILC map \citep{Bennett2003b,
Hinshaw2006}.  For $l<20$, the $a_{lm}$'s are taken from the ILC map
with no modification. At smaller scales, the ILC map is smoothed, so
a Gaussian random field is added such that the ensemble average
power equals the theoretical power spectrum taken from the WMAP
LAMBDA website (http://lambda.gsfc.nasa.gov/product/map/dr2/
params/lcdm\_all.cfm). We generate a map with HEALPix parameter
$N_{\rm side} = 4096$ (totaling 201,326,592 pixels).  This is then
interpolated onto the cylindrical coordinate system using the
bilinear interpolation subroutine included in the HEALPix
distribution \citep{Gorski2005}.

The point sources are drawn from prescribed number counts.  For a
given flux bin of width $\Delta S$, the number of sources in the
survey is drawn from a Poisson distribution with mean $  | dN(>S)/
dS | \Delta S \Delta \Omega$.  Within the bin, each source is
assigned a random flux.  We choose logarithmic bins that are narrow
compared to the range of source fluxes. Each set of sources has flux
counts tallied at some reference frequency $\nu_0$.  We scale the
source flux from this frequency using a power law, $S \propto
(\nu/\nu_0)^\alpha$.  Here $\alpha$ is a random variable, chosen for
each source from a Gaussian distribution with mean $\bar \alpha$ and
variance $\sigma^2_\alpha$ unique to each source population. For
radio sources we use \citet{Knox2004} number counts and choose $\bar
\alpha = -0.3$ and $\sigma_\alpha = 0.3$ for scaling parameters.
WMAP found a flatter spectrum with similar variance, but counts are
expected to steepen above 100 GHz. For infrared sources we use
\citet{Borys2003} number counts and choose $\bar \alpha = 3$ and
$\sigma_\alpha = 0.5$.

For the contribution of galactic thermal dust emission, we use the
``model 8'' prediction from \cite{Finkbeiner1999}, an extrapolation
to microwave frequencies of the dust maps of \cite{Schlegel1998}.
This model is a two-component fit to IRAS, DIRBE, and FIRAS data,
and is shown by \cite{Bennett2003} to be a reasonable template for
dust emission in the WMAP maps.  We create these galactic dust
emission maps for completeness of the simulations.  However, we do
not currently include these dust maps when we investigate cluster
detection.

\section{Y-M SCALING RELATION}

We would like to better understand the relation between SZ flux and
mass because this relation reflects cluster physics and also
provides the link between SZ cluster catalogs and cosmological
parameters, which are constrained by the cluster distribution as a
function of mass.  A more informed understanding about how changes
in cluster physics and cosmology alter the $Y-M$ relation will allow
for more information about cosmological parameters to be derived
from cluster surveys.

To investigate the SZ flux - mass cluster scaling relation, the
integrated Compton-y parameter is compared to cluster mass for the
clusters in our simulation.  The integrated Compton-y parameter is
the Compton-y parameter integrated over the face of the cluster,
i.e. $Y=(\int y d\Omega) d_{A}^{2}(z)$, where $d_{A}^{2}(z)$ is the
angular diameter distance.  In the self-similar model, a virialized
halo of mass $M_{\mathrm{vir}}$ has a virial temperature equal to
\begin{eqnarray}
T_{\mathrm{vir}}\propto [M_{\mathrm{vir}} E(z)]^{2/3},
\end{eqnarray}
where for a flat $\Lambda$CDM cosmology
\begin{eqnarray}
E(z)=[\Omega_{m}(1+z)^3 + \Omega_{\Lambda}]^{1/2}.
\end{eqnarray}
If clusters were isothermal, we would expect them to satisfy the
relation
\begin{eqnarray}
Y \propto f_{\mathrm{gas}} M_{\mathrm{halo}} T,
\end{eqnarray}
where $f_{\mathrm{gas}}$ is the cluster gas mass fraction. Thus we
find
\begin{eqnarray}
Y\propto f_{\mathrm{gas}} M_{\mathrm{vir}}^{5/3} E(z)^{2/3},
\label{eq:self-sim}
\end{eqnarray}
for the self-similar SZ flux - mass scaling relation. Since clusters
are not isothermal and not always in virial equilibrium, deviations
from the self-similar relation are expected. Moreover, variation of
$f_{\mathrm{gas}}$ with redshift and cluster processes such as star
formation and feedback can also cause deviations from
self-similarity.

\begin{figure}
\begin{center}
\epsscale{1.17}\plotone{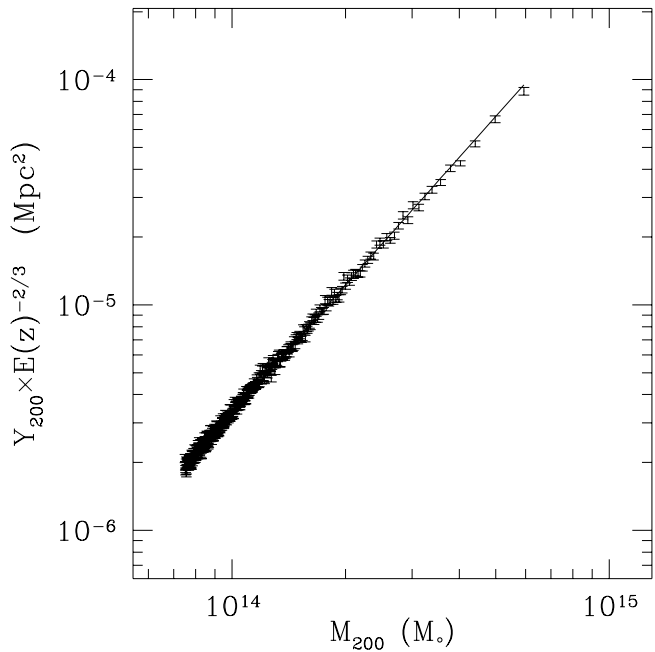} \caption{$Y_{200}\times E(z)^{-2/3}$
vs. $M_{200}$ for $\approx 10^5$ clusters within the N-body plus gas
simulation. Each point represents the mean value for 250 clusters,
and error bars represent the error on the mean.  The parameters and
reduced $\chi^{2}$ for the best-fit line are given in Table
\ref{tab:y_m}.} \label{fig:y_m}
\end{center}
\end{figure}

From the cluster catalog described in \S 2, we choose a sub-sample
with $M_{200} > 7.5 \times 10^{13} M_{\odot}$ to stay well above our
catalog mass limit of $5 \times 10^{13} M_{\odot}$ and ensure
catalog completeness. This leaves us with a sample of about $10^5$
clusters. For each cluster, we calculate $Y_{200}$, which is the
projected SZ Compton-y parameter in a disk of radius $R_{200}$.
Clusters are then ranked by mass, and for bins of 250 clusters each,
the mean of $M_{200}$ as well as the mean and standard deviation on
the mean of $Y_{200} \times E(z)^{-2/3}$ are calculated in each bin.
These values are plotted in Figure \ref{fig:y_m}. We then fit these
values to the power-law relation
\begin{eqnarray}
\frac{Y_{200}}{E(z)^{2/3}} = 10^{\beta} \Big( \frac{M_{200}}{10^{14}
M_{\odot}} \Big)^{\alpha}. \label{eq:y_m}
\end{eqnarray}
We find $\alpha=1.876 \pm 0.005$ and $\beta=-5.4774 \pm 0.0009$ with
a reduced $\chi^{2}$ of 1.004. This suggests that the $Y-M$ relation
is close to a power-law. This slope is steeper than that for the
self-similar model given in equation \ref{eq:self-sim}. This
steepening is expected since we assume feedback is independent of
total cluster mass.  Energy input of roughly 3 keV per particle will
clearly have a relatively larger effect on small clusters with lower
virial velocities than on more massive ones. The feedback reduces
the gas mass fraction more in lower mass clusters, as more gas is
pushed out in proportion to the cluster's total mass (i.e., the gas
fraction inside $R_{200}$ is smaller).  As a result, $Y_{200}$ is
also decreased more in the lower mass clusters.  The power-law index
that we find is slightly steeper than that found by some
hydrodynamic simulations (e.g.
\citet{Nagai2006,Motl2005,White2002b}).  This is understandable
since we include feedback from active galactic nuclei as well as
supernovae, which is different from the hydrodynamic simulations,
and the increased level of feedback steepens the slope. It has been
estimated that the feedback from active galactic nuclei may be
roughly an order of magnitude larger than that from supernovae
\citep{Ostriker2005}, so it potentially has a significant effect.
The slope we find is also in between that found by
\citet{daSilva2004} for their hydrodynamic simulations including
radiative cooling alone and cooling plus preheating.

A departure from a power-law is observed when we restrict our
cluster sample to have higher minimum masses.  As the minimum mass
of the sample is increased, the power-law index gets flatter,
illustrating the important effect of feedback on low-mass clusters
as well as the fact that the lowest mass clusters dominate the slope
of the $Y-M$ relation because their numbers increase rapidly as mass
is decreased.  We illustrate this departure in Table 2 where we have
preformed a similar analysis as above for clusters with $M_{200} > 2
\times 10^{14} M_{\odot}$, this time using bins of 20 clusters each.
We find $\alpha=1.81 \pm 0.02$ and $\beta=-5.463 \pm 0.009$ with a
reduced $\chi^{2}$ of 1.14. This departure indicates that while the
$Y-M$ relation is close to a power-law, there is some curvature.

\begin{figure}
\begin{center}
\epsscale{1.17}\plotone{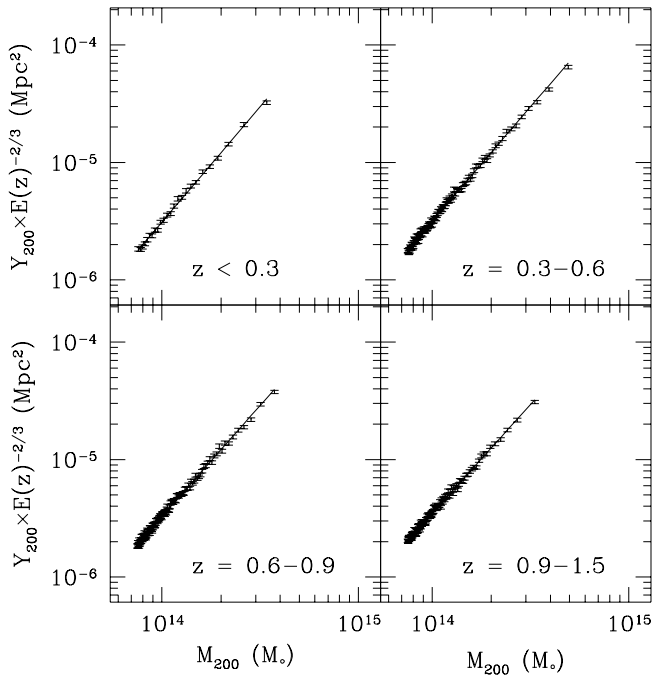} \caption{$Y_{200}\times E(z)^{-2/3}$
vs. $M_{200}$ for four different redshift regions. As in Figure 5,
each point represents the mean of 250 clusters, and error bars
represent the error on the mean. The parameters and reduced
$\chi^{2}$ for the best-fit lines are given in Table \ref{tab:y_m}.}
\label{fig:y_m2}
\end{center}
\end{figure}

We also divide our clusters above $M_{200} > 0.75 \times 10^{14}
M_{\odot}$ into four different redshift bins $(z<0.3, z\in(0.3,0.6),
z\in(0.6,0.9)$, and $z\in(0.9,1.5))$ to see the evolution of the
power-law index and normalization with redshift. The best-fit lines
to $Y_{200} \times E(z)^{-2/3}$ versus $M_{200}$ are shown in Figure
\ref{fig:y_m2}, and the best fit $\alpha$ and $\beta$ are given in
Table \ref{tab:y_m}. Some slight evolution of the power-law index
and normalization with redshift is apparent.  This is expected since
clusters are not perfectly self-similar. Furthermore, the stellar to
gas mass ratio is a function of redshift in our simulations, and
thus the feedback, which is proportional to the star formation, is
also a function of redshift. Since more feedback lowers $Y$
disproportionately at the low-mass end, we find the steepest slope
for the clusters with $z<0.3$, which have undergone more star
formation and feedback.  We find the flattest slope for the clusters
in the highest redshift bin.  Examining only the higher mass
clusters, $M_{200} > 2 \times 10^{14} M_{\odot}$, we see little
obvious evolution with redshift of the slope and normalization
considering their respective error bars (cf. Table 2). This again
suggests that the higher mass clusters are less sensitive to
feedback processes.

The scatter that we find for this $Y-M$ relation is most likely
underestimated, as we assume that the gas is described by
hydrostatic equilibrium, a single constant polytropic index, and the
same amount of feedback per stellar mass for all clusters. Also, we
have not included other sources of non-thermal pressure support
which may contribute non-negligibly to the total pressure.  Some of
these effects, such as turbulent pressure (see \citet{Rasia2004}),
are included in all current high-resolution hydrodynamic simulations
of galaxy clusters. Initial attempts have also been made to include
such effects as magnetic fields (e.g. \citet{Dolag1999}) and cosmic
rays \citep{Pfrommer2006}. Hydrodynamic simulations can provide a
robust check and calibration for semi-analytic models, but
continuing work is required to include all of the relevant gas
physics in both methods.  Possible changes in the parameters of the
predicted $Y-M$ relation can be expected as models achieve greater
realism.

\begin{table*}
\begin{center}
\begin{tabular}{|c|c|c|c|c|c|c|c|}
\hline M$_{200}>0.75 \times 10^{14}$ M$_{\odot}$&$\alpha$&$\sigma_{\alpha}$&$\beta$&$\sigma_{\beta}$&reduced $\chi^{2}$&clusters&bins \\
\hline
\hline z $>0$&1.876&0.005&-5.4774&0.0009&1.004&90750&363\\
\hline z $<0.3$&1.96&0.02&-5.509&0.004&1.18&6250&25\\
\hline z $=0.3-0.6$&1.944&0.009&-5.502&0.002&1.38&24750&99\\
\hline z $=0.6-0.9$&1.88&0.01&-5.484&0.002&1.14&27250&109\\
\hline z $=0.9-1.5$&1.85&0.01&-5.457&0.002&1.02&27500&110\\
\hline z $>1.5$&1.82&0.05&-5.421&0.004&0.66&4250&17\\
\hline
\end{tabular}
\caption{Best-fit parameters for the best-fit lines depicted in
Figures \ref{fig:y_m} and \ref{fig:y_m2}, fit using the power-law
given in equation \ref{eq:y_m}. \label{tab:y_m}}
\end{center}
\end{table*}

\begin{table*}
\begin{center}
\begin{tabular}{|c|c|c|c|c|c|c|c|}
\hline M$_{200}>2 \times 10^{14}$ M$_{\odot}$&$\alpha$&$\sigma_{\alpha}$&$\beta$&$\sigma_{\beta}$&reduced $\chi^{2}$&clusters&bins \\
\hline
\hline z $>0$&1.81&0.02&-5.463&0.009&1.14&7960&398\\
\hline z $<0.3$&1.85&0.05&-5.48&0.02&1.05&940&47\\
\hline z $=0.3-0.6$&1.78&0.03&-5.44&0.01&1.11&3080&154\\
\hline z $=0.6-0.9$&1.85&0.04&-5.49&0.02&1.48&2420&121\\
\hline z $=0.9-1.5$&1.82&0.05&-5.46&0.02&0.78&1400&70\\
\hline
\end{tabular}
\caption{Best-fit parameters for clusters with M$_{200}>2 \times
10^{14}$ M$_{\odot}$, fit using the power-law given in equation
\ref{eq:y_m}. \label{tab:y_m2}}
\end{center}
\end{table*}

\section{CLUSTER DETECTION}

To study cluster detection, three maps of the ACT strip are first
created (combining the primary microwave background, infrared and
radio point sources, and the full SZ effect) at the frequency
channels of 145, 215, and 280 GHz.  The ACT instrument is simulated
by convolving the ACT strip maps at 145, 215, and 280 GHz with a
Gaussian beam of full-width at half-maximum equal to 1.7', 1.1', and
0.93' respectively.  Gaussian random noise is then added to the maps
of $\sigma$ equal to $2\mu$K, $3.3\mu$K, and $4.7\mu$K per beam for
the three frequencies respectively. The above are preliminary
detector noise estimates for extensive dedicated observations of the
ACT strip. We produce a single map of the SZ clusters by applying a
multi-frequency Wiener filter to the three ACT maps. This filter
requires a model power spectrum of the first-order thermal SZ signal
from the clusters (see Eqn. \ref{eqn:thsz}). We first use a spectrum
derived from the simulations themselves, and then one derived from
the analytic prescription given by \citet{Komatsu2002} to test the
sensitivity of the filter performance on the input cluster model.
Clusters are then identified within the filtered map using a simple
peak-finding algorithm. These clusters are matched to a catalog
documenting the input map. From this matching process, we determine
the completeness and purity of our detected cluster sample.

\subsection{Multi-Frequency Wiener Filter}

In the literature, a number of different filters have been employed
to extract the SZ signal from microwave simulations (see
introduction for references).  Two commonly employed filters are
matched or scale-adaptive filters and Wiener filters.  Matched
filters make an explicit assumption about cluster profiles while
Wiener filters assume knowledge of the SZ power spectrum.  Both have
qualitatively similar shapes: they suppress the primary microwave
background at large scales and noise at small scales.  Our
motivation for choosing to use a Wiener filter stems from the fact
that it recovers the minimum variance SZ map and is simple to
implement.  We compare our results with \citet{Melin2006} who employ
a scale-adaptive filter and find a Wiener filter yields comparable
results.  We also use two different Wiener filters and find the
differences are minor.

Below we describe the map filter in detail. We wish to recover one
filtered map (of clusters) from several ACT maps (at three different
frequencies).  The latter are described by the data vector
\mbox{\boldmath$d$}. We write \mbox{\boldmath$d$} =
$\mathrm{R}$\mbox{\boldmath$s$} + \mbox{\boldmath$n$}, where
$\mathrm{R}$ is the instrument response to the Compton-y signal map
\mbox{\boldmath$s$} (see Eqn. \ref{eqn:Compton-y}), and
\mbox{\boldmath$n$} is the noise. In this notation, $\mathrm{R}$
includes both the frequency dependence of the first-order thermal SZ
signal (see Eqn. \ref{eqn:thsz}) and the convolution with the beam.
The noise \mbox{\boldmath$n$} includes detector noise, primary
microwave background, point sources, and contributions to the SZ
signal other than from the first-order thermal SZ. We denote the
covariance of the signal, noise, and data by $\mathrm{S}=\langle
\mbox{\boldmath$s$}\mbox{\boldmath$s$}^{\dagger}\rangle $,
$\mathrm{N}=\langle
\mbox{\boldmath$n$}\mbox{\boldmath$n$}^{\dagger}\rangle $, and
$\mathrm{D}=\langle
\mbox{\boldmath$d$}\mbox{\boldmath$d$}^{\dagger}\rangle =
\mathrm{RSR^\dagger+N}$, assuming signal and noise are largely
uncorrelated.

The multi-frequency Wiener filter is given by \begin{eqnarray}
\mathrm{W}=\mathrm{S}\mathrm{R}^\dagger\left[\mathrm{R}\mathrm{S}\mathrm{R}^\dagger+\mathrm{N}
\right]^{-1} = \mathrm{S}\mathrm{R}^\dagger\mathrm{D}^{-1},
\end{eqnarray} and it gives the least-squares signal map reconstruction
via ${\hat{\mbox{\boldmath$s$}}} = \mathrm{W} \mbox{\boldmath$d$}$.
Here we are assuming statistically homogeneous signal and noise, so
all the covariances are (block) diagonal in the Fourier
representation: they are the auto and (frequency) cross-spectra.
This makes the application of the Wiener filter straightforward:  we
Fourier transform the ACT maps, apply the filter, and inverse
Fourier transform to obtain the SZ cluster map. (See
\citet{Tegmark1996} for a useful reference on Wiener filtering.)

\subsection{Filtering Maps}

To filter the maps, we divide the ACT strip (with SZ, primary
microwave background, and point sources combined) into roughly
$4^{\circ} \times 3.25^{\circ}$ patches.  The above dimensions are
chosen to give square patches in pixel space.  To compute
$\mathrm{D}$, the power and cross-power spectra for each patch is
calculated and averaged over the whole strip.  For the first model
cluster power spectrum ($\mathrm{S}$), we compute the spectrum from
the simulations in a similar fashion.  Later we also use the
semi-analytic spectrum of \citet{Komatsu2002}.

To avoid aliasing of signal at the edges of each patch, we use an
overlap-and-save method to filter overlapping pieces of the map.
This permits us to discard all the vertical edges (edges of constant
right ascension) on the patch borders \citep{Press1997}.  In this
way, we remove any discontinuities in the filtered map at the
boundaries of the patches. The horizontal edges are less of a
concern since they can be discarded after the whole strip is
filtered.  This filtering procedure provides a filtered map of the
entire ACT strip consisting of the recovered SZ signal.

\subsection{Identifying Clusters}

Clusters are identified by searching for all peaks in the filtered
map above a given threshold value.  The threshold cuts we apply are
1, 2, and 3 times the standard deviation of the filtered map.  A
peak is identified simply as a pixel larger than any pixel within a
radius $r=6p$ around it, where $p$ is the pixel size ($p\approx
0.21'$ for these maps).  We choose this radius both because $\approx
1.3'$ is a typical cluster size and in an effort to be consistent
with \citet{Melin2006} for comparison purposes. Once a list of peaks
is compiled, the peaks are matched to clusters in the input cluster
catalog. We identify a match if the peak and catalog cluster are
within a distance r of one another. Multiple peaks matching a single
catalog cluster are allowed, as are multiple catalog clusters
matching a single peak. Follow-up observations to determine cluster
redshifts should sort these cases out.  Any peak that is not within
$r$ of any catalog cluster, is flagged as a false detection.

It is possible that some peaks match catalog clusters just by chance
alignment.  However, since clusters down to only $5\times 10^{13}
M_{\odot}$ are included, all the clusters in the simulation are
comparable to or above ACT instrument noise, and thus are in
principle detectable. So we accept that some small fraction of the
simulation clusters may be detected by chance alignment with
spurious peaks and do not try to distinguish these cases from real
matches.

\subsection{Results}

\begin{figure*}
\begin{center}
\epsscale{1.0}\plotone{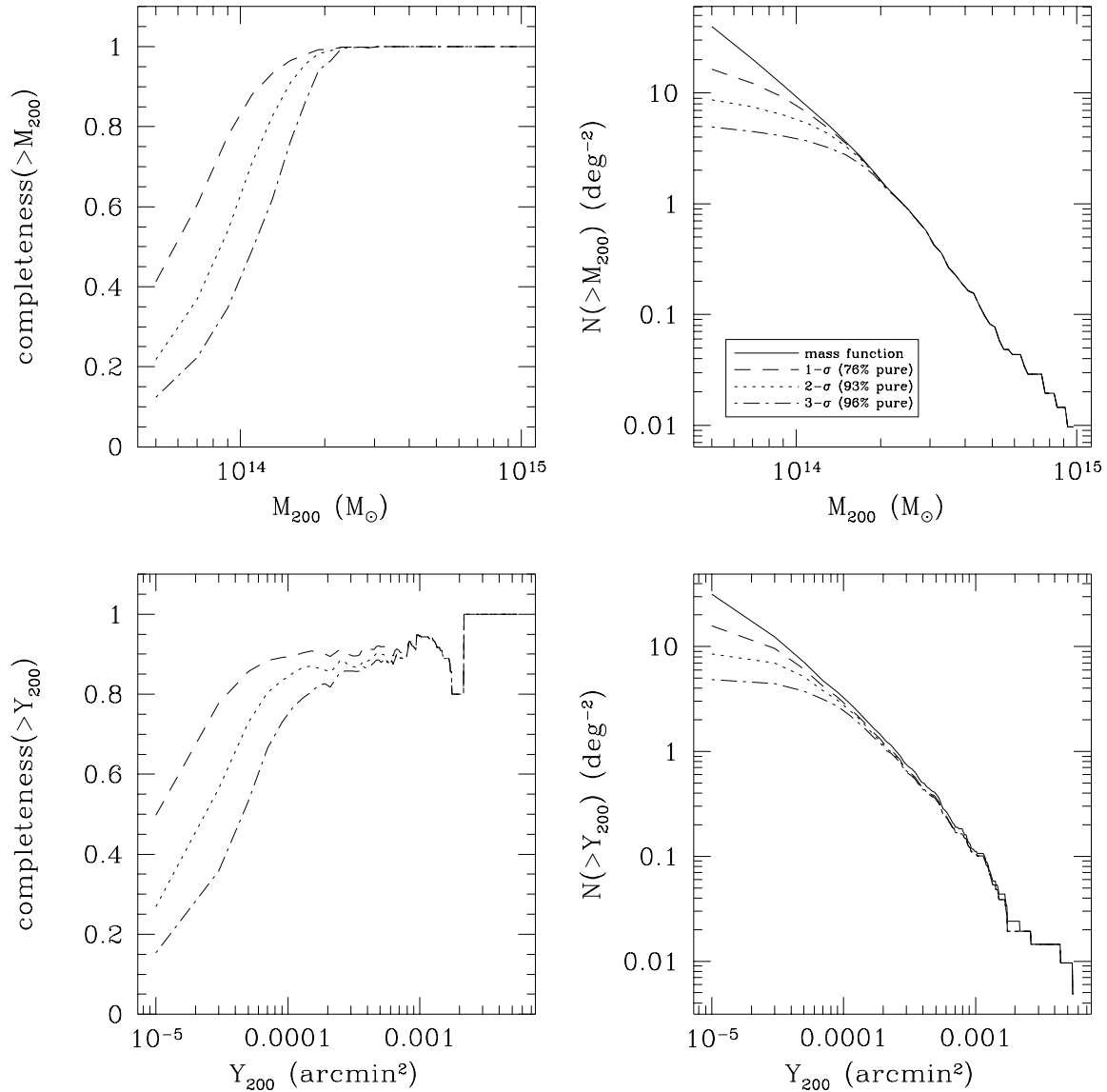} \caption{ Completeness and purity of
the detected cluster sample with ACT instrument specifications, in
the case of no point sources. Different curves represent 1-$\sigma$,
2-$\sigma$, and 3-$\sigma$ threshold cuts in the detection
algorithm, where $\sigma$ is the standard deviation of the filtered
map.  The solid curve represents the simulation mass function.  The
purity of each cluster sample is given in the legend.  See text for
further details.} \label{fig:nopoints}
\end{center}
\end{figure*}

First we examine the case when the three-frequency ACT strip maps
contain no radio or infrared point sources.  The completeness and
number of detected clusters per deg$^{2}$ are given in Figure
\ref{fig:nopoints} as a function of both $M_{200}$ and $Y_{200}$
(the latter measured as a solid angle in units of arcmin$^{2}$). The
dashed, dotted, and dash-dotted curves represent the 1-$\sigma$,
2-$\sigma$, and 3-$\sigma$ threshold cuts respectively, and the
solid line represents the simulation mass function.  The purity of
each cluster sample is given in the legend (purity $=1-\%$ of
false-positive detections; completeness = number of detected
clusters/number of input clusters).  In the absence of point
sources, using a 3-$\sigma$ threshold cut, our sample of detected
clusters is 96$\%$ complete down to $2 \times 10^{14} M_{\odot}$
with only $4\%$ contamination from false positives. The completeness
as a function of $Y_{200}$ and purity are comparable to that found
by \citet{Melin2006} when they investigated a no-point-source case
for the South Pole Telescope (SPT) using a matched filter and
simulated clusters that perfectly matched their filter.

In the completeness versus $Y_{200}$ plot, there is a sharp decrease
in the completeness from 1 to 0.8 at $Y_{200}=2\times 10^{3}$
arcmin$^{2}$.  This effect is caused by a single cluster at
$z\approx 0.04$ with $M_{200}\approx 8\times 10^{13} M_{\odot}$.
Since this cluster is so nearby, it extends over many pixels, and
thus has a large $Y_{200}$ in units of arcmin$^{2}$.  However, its
signal in each pixel is below that of the 1-$\sigma$ threshold cut,
so it is not detected.  If the plot had been made with $Y_{200}$ in
units of physical area (Mpc$^{2}$), this feature would disappear.
Therefore, this feature is an artifact reflecting the fact that
arcmin$^{2}$ is not a unit intrinsic to a cluster.  It also suggests
that for detecting extended, low-redshift clusters, searching for a
peak pixel is not the best method.  Profile-fitting may provide a
better alternative in this case.

\begin{figure*}
\begin{center}
\epsscale{1.0}\plotone{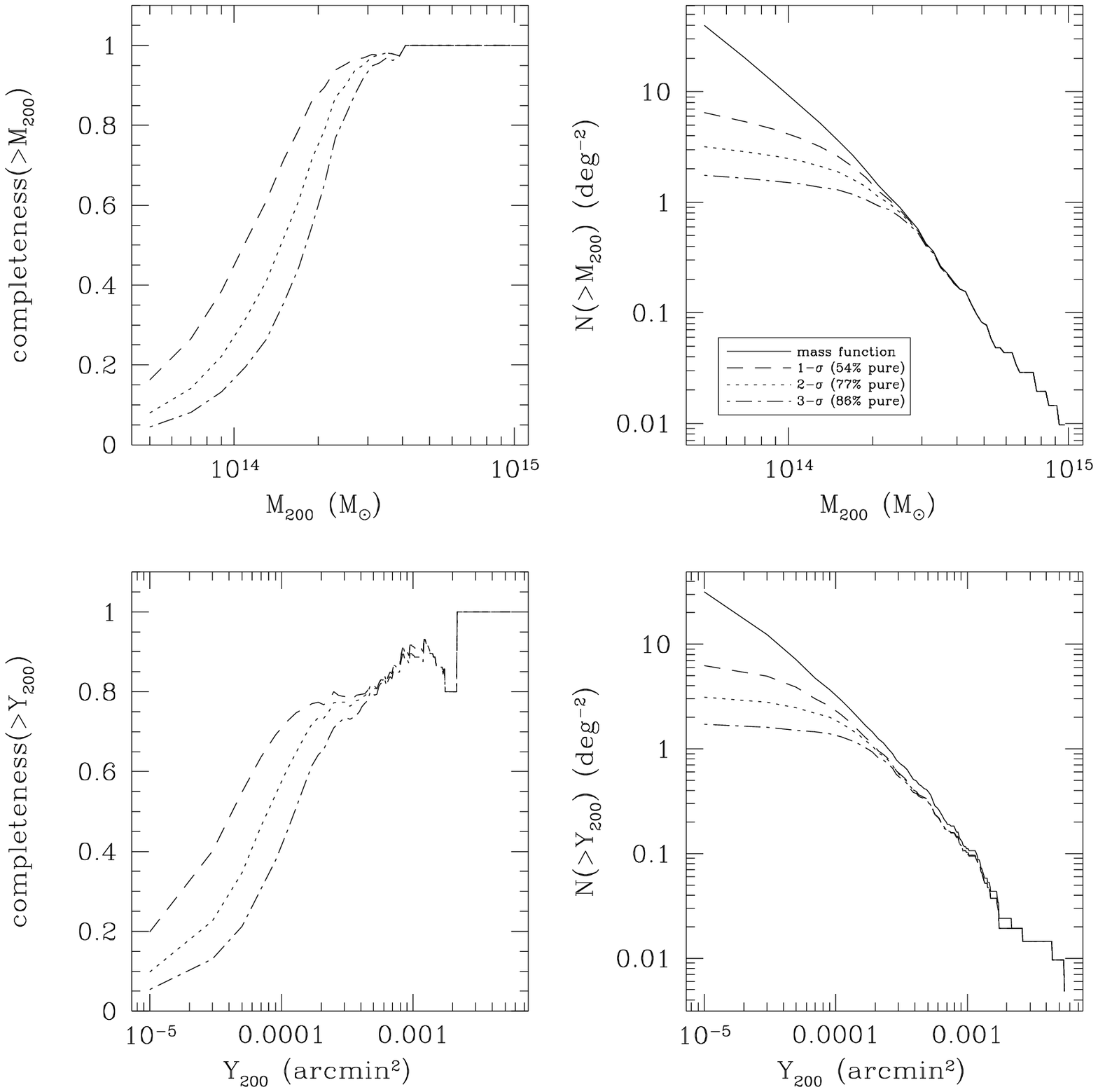} \caption{Completeness and purity of
the detected cluster sample in the case of infrared point sources
only.} \label{fig:ironly}
\end{center}
\end{figure*}

Next, infrared point sources are added to the three-frequency ACT
maps following the prescription described in \S 2.3.  Infrared
sources up to a flux limit of $\approx 0.4$ Jy at 145 GHz are
included since the number of sources with higher flux is effectively
zero for the area of the ACT strip given the above prescription.
Figure \ref{fig:ironly} shows the cluster detection results after
adding this population of point sources.  For this case, using a
3-$\sigma$ threshold cut, the detected cluster sample is 95$\%$
complete down to $3\times 10^{14} M_{\odot}$ with 14$\%$
contamination from false-positive detections. The percentage of
false-positives is comparable to that found by \citet{Melin2006}
when they used N-body, as opposed to spherical, clusters for their
no-point-source case.  A direct comparison of completeness,
including infrared point sources, with their work is difficult
because we do not use isothermal, spherical clusters and we include
clusters down to a lower mass limit, which crowds the field and
increases confusion.

\begin{figure*}
\begin{center}
\epsscale{1.0}\plotone{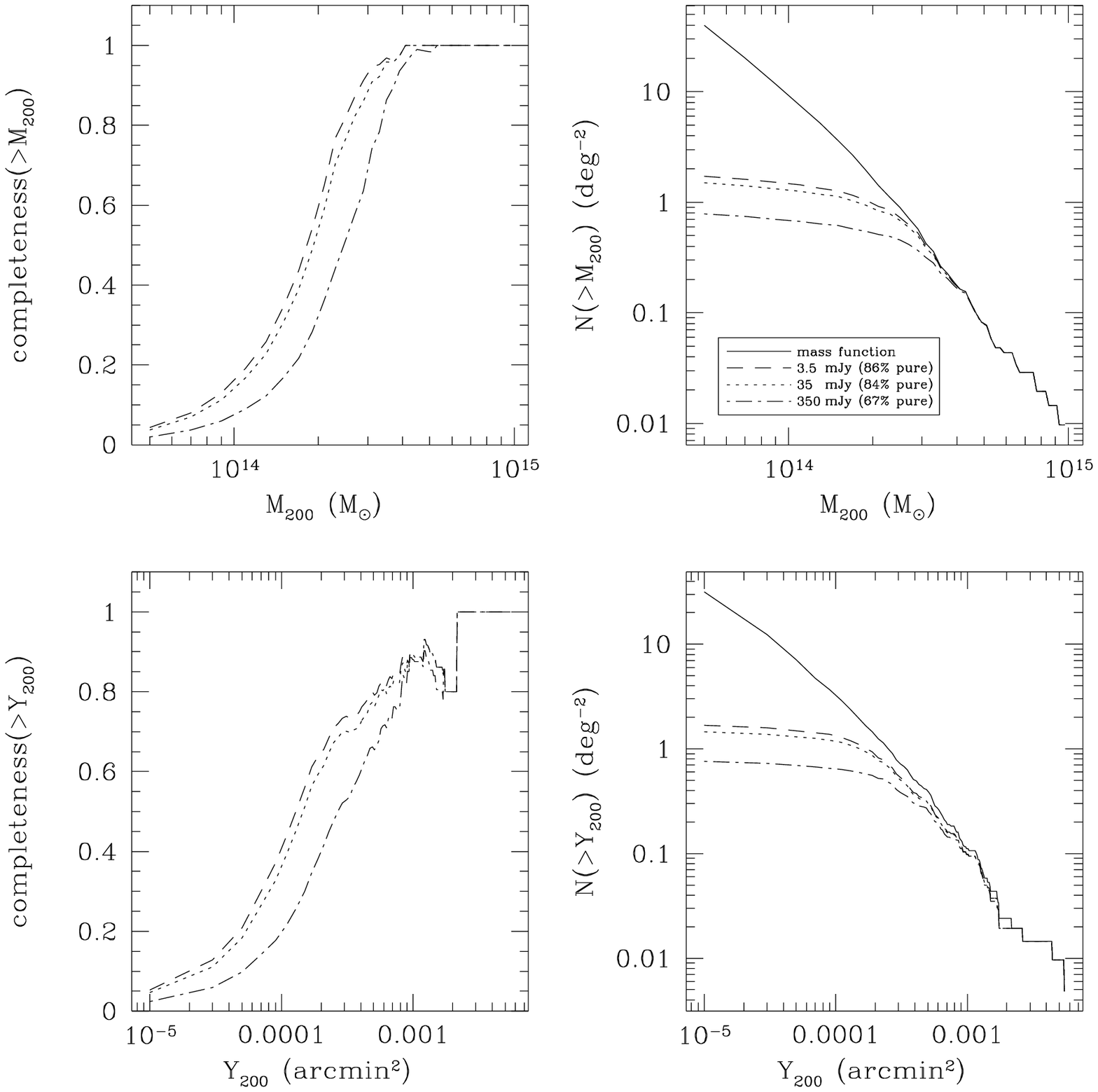} \caption{Completeness and purity when
both infrared and radio sources are included.  A 3-$\sigma$ cut is
used for all the curves. The different curves represent different
flux cuts at 145 GHz for the radio sources. If all the radio sources
above 3.5 mJy at 145 GHz are removed from the three-frequency ACT
maps, then we recover the results with only infrared point sources
included. Removing all sources above 35 mJy at 145 GHz gives results
that are only slightly degraded compared to the former case.}
\label{fig:irandrad}
\end{center}
\end{figure*}

We further investigate the case where both infrared and radio point
sources are added to the three-frequency ACT strip maps.  The radio
sources follow the prescription described in \S 2.3.  Figure
\ref{fig:irandrad} shows the cluster detection results including
radio sources in the ACT maps below three different flux limits (3.5
mJy, 35 mJy, and 350 mJy at 145 GHz).  All the curves correspond to
cluster detection using a 3-$\sigma$ threshold cut.  The dashed
curve shows that if all the radio sources above 3.5 mJy at 145 GHz
are removed from the three-frequency ACT maps, the
infrared-point-source-only case is recovered, depicted in Figure
\ref{fig:ironly}. If all the radio sources above 35 mJy at 145 GHz
are removed from the three ACT maps, we obtain results that are of
only slightly lower quality than the previous case.  These results
can be understood by noting that the inclusion of radio sources
below 35 mJy and infrared sources increases the average power in the
simulated ACT strip on scales of a few arcminutes by roughly an
order of magnitude, as compared to not including point sources.  The
inclusion of all the radio sources above 35 mJy increases the power
on these scales by several orders of magnitude more.  This large
increase in noise power is why these few very bright sources limit
cluster detection with this technique. For an arcmin$^{2}$ beam at
145 GHz, 3.5 mJy and 35 mJy correspond to $\approx 100\mu$K and
$\approx 1000\mu$K respectively. Thus achieving the removal of radio
sources above the former flux limit may require interferometric
observations of the ACT strip at a lower frequency (e.g. with the
Australia Telescope Compact Array (ATCA) at $\approx$ 20 GHz).
Achieving the latter flux limit of radio sources is most likely
possible with ACT alone. (Sources of $\approx 1000\mu$K should be
easy to identify, and we find $\approx 700$ of them in our
simulations of the ACT strip, which is not an unmanageable number to
remove.)  Assuming removal of radio sources above 35 mJy at 145 GHz,
our detected cluster sample is 90$\%$ complete down to $3 \times
10^{14} M_{\odot}$ and $84\%$ pure.

\begin{figure*}
\begin{center}
\epsscale{1.0}\plotone{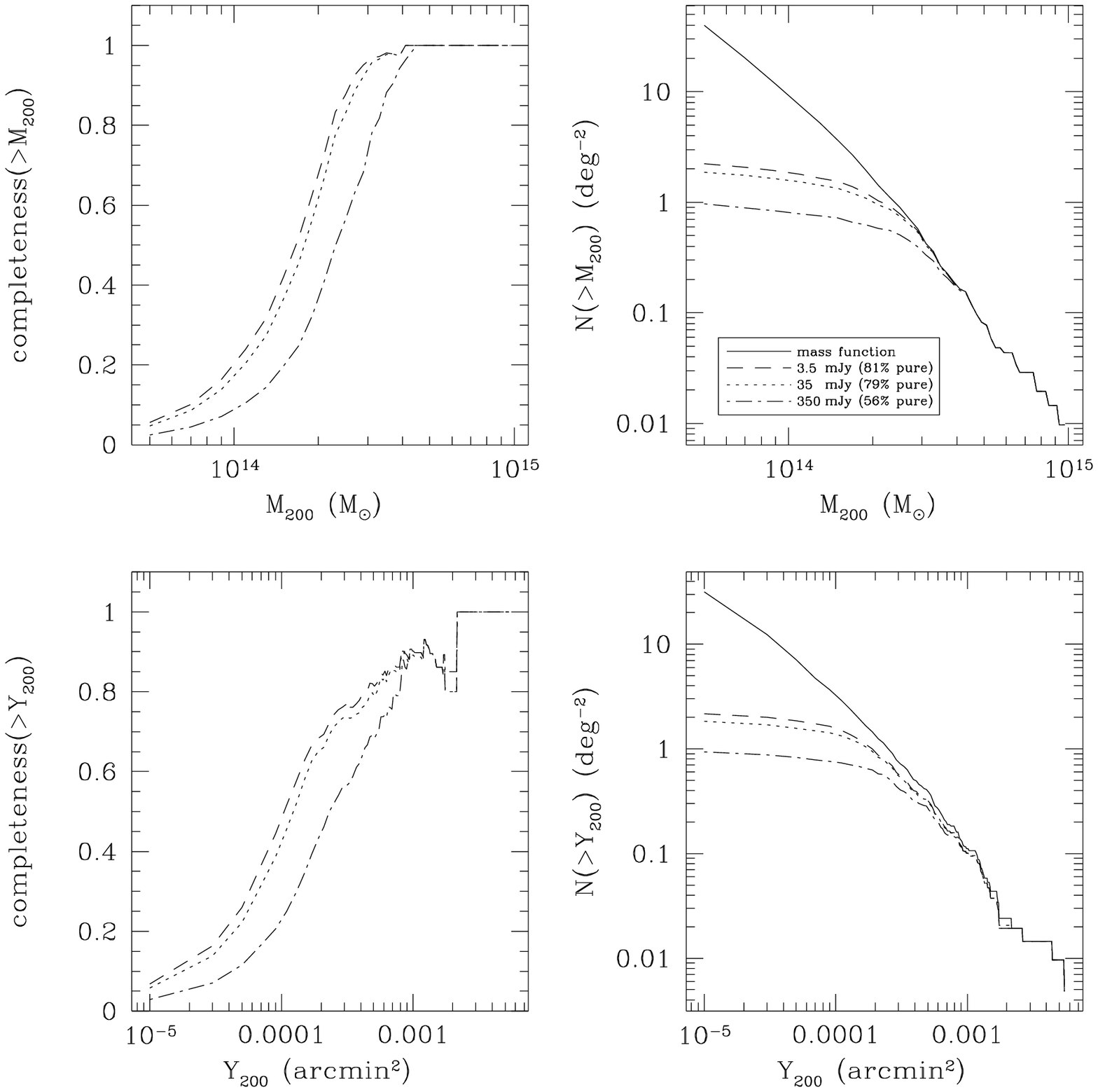} \caption{Completeness and purity of
the detected cluster sample for the cases described in Figure
\ref{fig:irandrad}, but this time using a filter constructed with a
cluster Compton-y power spectrum derived analytically from
\citet{Komatsu2002}, instead of taken directly from the simulations.
The completeness is slightly increased and the purity slightly
decreased as compared to the previous Figure, but overall the
difference is small.} \label{fig:irandrad_KS}
\end{center}
\end{figure*}

The filter used in obtaining the above results was created using the
Compton-y power spectrum from the simulations themselves.  Since we
shall not know beforehand the true cluster power spectrum, we remake
the filter using a Compton-y power spectrum derived analytically
from the prescription in \citet{Komatsu2002}. With this filter, we
repeat the above exercise.  Figure \ref{fig:irandrad_KS} displays
the cluster detection results using this filter and adding both
infrared and radio point sources.  The different curves again
correspond to the three different flux cuts of radio sources, and a
3-$\sigma$ threshold cut is used for them all.  For the three flux
cut cases, there is a slight increase in completeness of the cluster
sample as compared to Figure \ref{fig:irandrad}. However, there is
also a slight decrease in the purity of the cluster sample.  If we
compare the first-order thermal SZ power spectrum from the
simulations and derived analytically (Fig. \ref{fig:thsz}), we see
that the filter made with the analytic power spectrum leaves more
small scale power in the filtered map. This increases the sample
completeness, but also decreases its purity, as more point sources
are left in the map as well. Overall, however, the difference in
using the analytic power spectrum in the filter, as compared to the
power spectrum from the simulations, is small. This suggests that
the performance of the filter is not overly sensitive to how the
cluster physics is modeled when constructing it.

\section{FUTURE WORK}

In this work, we did not correlate the radio and infrared point
sources with the clusters.  This was due to a lack of observational
data regarding these correlations.  If either population is both
strongly correlated with clusters and exhibits a relatively flat
spectrum between the infrared/radio regime and the microwave, then
some further percentage of clusters may fail to be detected.  Recent
BIMA and OVRO observations targeting massive galaxy clusters at 28.5
GHz suggest an overabundance of mJy radio sources in these massive
clusters as opposed to non-cluster fields \citep{Coble2006}.
However, they also find a steeper spectral index for these sources
as compared to the brighter, rarer population of radio sources found
by WMAP \citep{Bennett2003}.  If their spectral index steepens
further above 30 GHz, these former sources may not be an issue.  But
if their slope flattens out, some of these sources may fill in the
SZ decrement of some percentage of clusters at 145 GHz.  Without a
better understanding of these correlations and observations of these
point sources in the range of 145 to 280 GHz, it is difficult to
model both infrared and radio sources more realistically.  Upcoming
microwave surveys and follow-up or concurrent radio and infrared
observations overlapping the survey areas will hopefully clarify
these current unknowns.  It is useful to keep in mind that, for
constraining cosmology, obtaining a large number of detected
clusters is less important than obtaining a solid understanding of
the selection criteria.  So if point sources affect cluster
detection more significantly than anticipated, as long as this
effect can be accurately characterized, it should not pose an
overwhelming hurdle for cosmological investigations.

Some further avenues of investigation that we did not cover in this
work include the following.  The instrument noise can be more
realistically modeled than we have done here, and atmospheric noise
can be included.  The lensing of the primary microwave background
and point sources by clusters can be modeled to study its effect on
cluster detection and SZ flux recovery.  Moreover, the optimal
recovery of cluster SZ flux, which is an issue distinct from cluster
detection, should be investigated.  This will aid in flux - mass
identifications and SZ component separation.  Some recent work
addressing the later issue can be found in \citet{Melin2006}.  We
also by no means suggest that Wiener filtering is the ideal cluster
detection tool, and further study in this direction is also
warranted.

\section{CONCLUSION}

New microwave instruments have the potential to detect many galaxy
clusters out to redshift z=1 and beyond by their SZ signature. These
galaxy clusters provide information about the growth of structure,
which in turn constrains cosmology.  However, the SZ signal is
embedded within the signals from the primary microwave background,
radio and infrared point sources that still have significant flux in
the microwave regime, and galactic emission.  Thus, in an effort to
study the detection of clusters via their SZ effect, we simulated
the microwave-sky over both the proposed ACT observing region and a
parallel strip centered at $\delta=-5$ degrees. To realistically
model the small scale cluster physics, which affects the SZ signal,
we combined an N-body simulation with an analytic gas model
\citep{Ostriker2005, Bode2006}. This allowed us to model cluster
processes such as star formation and feedback over much larger
volumes than can currently be achieved by hydrodynamic simulations.
One product of these cluster simulations is that we are also able to
study features in the SZ flux - mass scaling relation, such as its
slope and normalization and their evolution with redshift. Knowledge
of this relation is necessary to connect SZ flux to cluster mass,
which is the quantity of interest for cosmology.

In investigating the relation between SZ flux and mass, roughly
$10^5$ clusters from our N-body plus gas simulations are fit to the
power-law relation given in equation \ref{eq:y_m}.  The best-fit
power-law index is found to be $\alpha=1.876 \pm 0.005$, and the
reduced $\chi^{2}$ to be 1.004.  This slope is steeper than 5/3,
which is the expectation for the self-similar model. This steeper
slope is consistent with the inclusion of feedback from both active
galactic nuclei and supernovae in these simulations which serves to
lower the SZ flux more for lower-mass clusters. This slope is
understandably steeper than that for hydrodynamic simulations, which
do not include feedback from active galactic nuclei and thus have a
lower amount of overall feedback than in this simulation.  We find
some redshift dependence for both the power-law index and
normalization of this relation. This is expected since clusters are
not in fact self-similar, and because star formation and cluster
feedback are redshift dependent in our simulations. The steepening
of the power-law index at lower redshifts is consistent with more
star formation and feedback occurring at later times.  Considering
only the higher mass clusters, $M_{200} > 2\times 10^{14} M\odot$,
we find a flatter slope of $\alpha=1.81 \pm 0.02$ with a reduced
$\chi^{2}$ of 1.14.  We also see less obvious evolution of the slope
and normalization with redshift.  This is consistent with the higher
mass clusters being less sensitive to feedback processes.  It should
be noted that the details of this $Y-M$ relation can change as
additional sources of non-thermal pressure support such as magnetic
fields, cosmic rays, and turbulence are included in semi-analytic
gas models.

Our projections for cluster detection with the ACT instrument are
also promising.  Cluster detection was investigated under varying
levels of point source contamination utilizing a multi-frequency
Wiener filter and peak-finding algorithm.  These results suggest
that in the absence of point sources, considering only peaks above
3-$\sigma$ in the filtered map, ACT can obtain a cluster catalog
that is 96\% complete above $2\times 10^{14} M_{\odot}$ and 96\%
pure (4\% false-detections).  The inclusion of infrared sources
results in a catalog that is 95\% complete above $3\times 10^{14}
M_{\odot}$, and 86\% pure.  When all radio sources are included in
addition, with no flux cutoff, we find that the noise from the
brightest radio sources interferes significantly with cluster
detection using this technique.  However, if the brightest sources
are removed from the ACT maps, there is a considerable improvement.
Removing all radio sources above 35 mJy at 145 GHz from the
three-frequency ACT maps results in a catalog that is 90\% complete
above $3\times 10^{14} M_{\odot}$ and 84\% pure.  Removing all radio
sources above 3.5 mJy at 145 GHz gives the same results as the
inclusion of only infrared point sources.  These results are
encouraging, since 35 mJy sources should be easily identifiable and
removable from ACT maps without need for further observations at
alternate frequencies.

This study was repeated using a cluster model in our filter that
differed from that in our simulations.  Doing this produced only a
small change in the cluster detection results. This suggests that
the filter performance is not very sensitive to the details of the
cluster model used to create it.  As a result, efforts to
characterize the selection function are made easier.

The potential of new microwave instruments to provide knowledge
about cluster physics and cosmology is substantial, as our results
further confirm.  It is hoped that the simulations created to pursue
the above studies will serve as a useful tool for future
investigations toward making this potential a reality.

\acknowledgments NS would like to thank Jack Hughes for many useful
discussions and comments.  The authors would also like to thank Jim
Bartlett, Gil Holder, Arthur Kosowsky, Robert Lupton, J.P. Ostriker,
Beth Reid, and David Spergel for useful discussions.  This work was
partially supported by the National Center for Supercomputing
Applications under grant MCA04N002. In addition, computational
facilities supported by NSF grants AST-0216105 and AST-0408698 were
used, as well as high performance computational facilities supported
by Princeton University under the auspices of the Princeton
Institute for Computational Science and Engineering (PICSciE) and
the Office of Information Technology (OIT).  NS acknowledges NSF
PIRE grant OISE-0530095 and NASA LTSA NAG5-11714.  HT was supported
in part by NASA LTSA-03-000-0090, and KH's portion of this work was
partially performed at the Jet Propulsion Laboratory, California
Institute of Technology, under a contract with the National
Aeronautics and Space Administration.

%\clearpage

%Bibliography-related commands
\bibliographystyle{apj}
%\bibliography{refs,paulsrefs}

\end{document}